\documentclass{iserc}% manuscript submitted for review
\usepackage{soul}
\usepackage{multirow}
\conference{Proceedings of the 2019 IISE Annual Conference\\
H.E. Romeijn, A. Schaefer, R. Thomas, eds.}% Do not change this line.

\title{\titlesize Interdependent Infrastructure System \\Risk \& Resilience to Natural Hazards
}
\author{
Benjamin Rachunok \\School of Industrial Engineering\\Purdue University, West Lafayette, IN, USA\\
\vspace{0.3cm}% Space beween authors with a different affiliation.
Roshanak Nateghi\\School of Industrial Engineering, Division of Environmental Engineering \\Purdue University, West Lafayette, IN, USA}
\authorlist{Rachunok, Nateghi}%Heading
\abstractID{583684}% Fill in the web conference management system-assigned abstract ID.

\begin{document}

\newcommand{\Perdos}{Paul Erd\H{o}s }
\newcommand{\er}{Erd\H{o}s R\`enyi }
\newcommand{\ie}{\emph{i.e. }}

\renewcommand*{\thefootnote}{\fnsymbol{footnote}}

\newpage
\maketitle

\begin{abstract}
%{\small All papers must include an Abstract. Begin with the word Abstract immediately following the title block with one blank line in between. Use centered, 12 point, Times New Roman Bold text for this heading. Use 10 point Times New Roman font for the text of the abstract. There should be a single blank line between the heading and this text. The abstract should be fully justified and consist of a single paragraph between 100 and 200 words (maximum of 200 words). The abstract must match the program abstract submitted in the abstract submission system}
Complex, interdependent systems are necessary to the delivery of goods and services critical to societal function. Here we demonstrate how interdependent systems respond to disruptions. Specifically, we change the spatial arrangement of a disruption in infrastructure and show that -while controlling for the size- changes in the spatial pattern of a disruption induce significant changes in the way interdependent systems fail and recover. This work demonstrates the potential to improve characterizations of hazard disruption to infrastructure by incorporating additional information about the impact of disruptions on interdependent systems.
\end{abstract}

\section*{Keywords}
Resilience, infrastructure, risk, networks

\section{Introduction}
Interdependence is inherent in many critical systems vital to the continuation of a nation's well-being \cite{ouyang2014}. Electricity, natural gas, transportation, and telecommunication are all provided by infrastructure systems which require bi-directional inter- and intra-system connection for optimal functionality. For example, telecommunication grids require continued power for operation, while the electric grid requires telecommunication networks to function \cite{booker2010,ouyang2014,winklerjames2011}. The criticality of the goods and services provided by these systems necessitates the design of resilient interdependent systems. In this work, we study the disruption and recovery of interdependent systems after a major disturbance and quantify the influence of changes in the spatial distribution of hazards on overall system resilience. 

Much attention has been given to the study of failures in interdependent networks from both a theoretical and applied perspective. Interdependence has been shown previously to improve overall system robustness to disruption \cite{duenas-osorio2009} at the expense of reducing steady-state performance. Previous work has shown that interdependent infrastructure systems will respond differently to an identical hazard or disruption due to their individual components and their topology (\emph{eg} telecommunications networks and water distributions will not be impacted similarly by a hurricane) \cite{duenas-osorio2009a}. Previous work has considered disruptions to the network which occur randomly \cite{duenas-osorio2009,erdener2014} -indicative of general system aging and degedration-  or via targeting \cite{hu2016} in which vertices are removed because of their importance. This work improves upon previous studies by considering the impact of changes to the \textit{spatial distribution} of failures on system performance while controlling for the influence of the size of the disruption. We hypothesize that --contrary to previous analyses-- the impact of hazards on interdependent systems does not follow a random pattern and may be clustered locally. To test our hypothesis, we change the spatial distribution of failures in each system and compare the resulting system performance immediately after failure and while the simulated systems are being repaired. This provides evidence to indicate that --when controlling for the size of the impact-- system performance is significantly influenced by the spatial distribution of outages. We further show that a significant change in system performance can be measured in both systems if disruptions are assumed to impact each system with a separate spatial distribution.

\section{Methods and Data}
To evaluate the impact of different outages on measurements of system performance, we simulate the failure an recovery of two interdependent systems in response to different sizes and spatial distributions of disruptions. The performance of each system is measured as it fails and is repaired. What follows is an overview of the simulation methodology, the calculation of performance metrics, and the data used to construct the systems.

\subsection{Methods}
Our analysis of the interdependent systems uses two graphs -representative of two infrastructure systems- and couples them to create interdependencies among the systems. The two networks, $g_1$ and $g_2$ are generated such that 

\[
g_1 = G(V_1,E_1) \qquad g_2 = G(V_2,E_2)
\]

each are made up an edge set, $E$, and a vertex set, $V$. The size of the edge set and vertex set  (\emph{i.e.} the number of edges and vertices) of $g_1$ are $|E_1|$, and $|V_1|$ respectively \cite{newman2005}. The degree of each vertex is the number of edges to which it connects, and here is represented as.  
\[
\text{Degree of vertex } ~i = d(v_i) , v_i\in V(g)
\] 

To generalize the connections between the vertices in opposing graphs, a dependence matrix $D_{g1,g2}$ is used to relate elements of $g_1$ to elements of $g_2$. $D_{g1,g2}$  is defined as a matrix of size $|V_1| \times |V_2|$. Elements of the matrix represent individual component-level dependencies. Consequently, $D_{g1,g2}(i,j)=1 $ if $v_j$ depends on $v_i$ to function and $v_i \in g_1$ and $v_j \in g_2$. This allows for the representation of directional dependence in failures and recovery. if $D_{g1,g2}(i,j)=D_{g1,g2}(j,i)$, then we have an interdependence between components, and if $D_{g1,g2}= D_{g1,g2}^t$ then the systems are fully coupled insofar as  $D_{g1,g2}(i,j)=D_{g1,g2}(j,i) ~\forall i \in \left[1,|V_1|\right] \text{ and } j \in \left[1,|V_2|\right]$. An example would be:

\[
D_{g1,g2}= 
\begin{bmatrix}
    0 & 1 & 0 & 0 &\dots \\
    0 & 1 & 0 & 0 &\dots \\
    1 & 1& 1 & 1 &\dots \\
    \vdots & \vdots & \vdots & \vdots & \ddots \\
\end{bmatrix}
\]

in which $v_1$ through $v_4$ in $g_2$ depend on $v_3$ to function. This provides a general framework to relate one network to the other in the failure and recovery of the systems. 

\subsection{Failures}
After $g_1$ and $g_2$ are coupled via $D$, outages are generated in each system. The methods of outage generation are discussed in subsequent sections. In the disruptions, the set of failures is comprised of two sets.  First are the failures directly induced by the disruption's impact on the system. Second is the dependent impacts within or across systems. The initial set of failures -\ie those induced by the hazard- in graph $g$ are denoted $f^f_g$ and the subsequent dependent failures in graph $g$ are denoted $f^d_g$. After the initial set of failures are generated in $g_1$ and added to $f^f_{g1}$, dependencies in $g_2$ are identified via $D$. $f^d_{g2}$ is updated to reflect the elements of $g_2$ which fail as a result of a failure in $g_1$. $f^d_{g1}$ is then updated based on $f^d_{g2}$. The failures cascade across the the two networks until no more dependencies are found. The process is repeated starting with $f_{g2}^f$ and propagates until equilibrium. The results of the failure generation represent the total, initial impact of a disruption on the interdependent system. 

\subsection{Failure Generation Methods}
\begin{figure}[htb]
\centering
\includegraphics[width=4in]{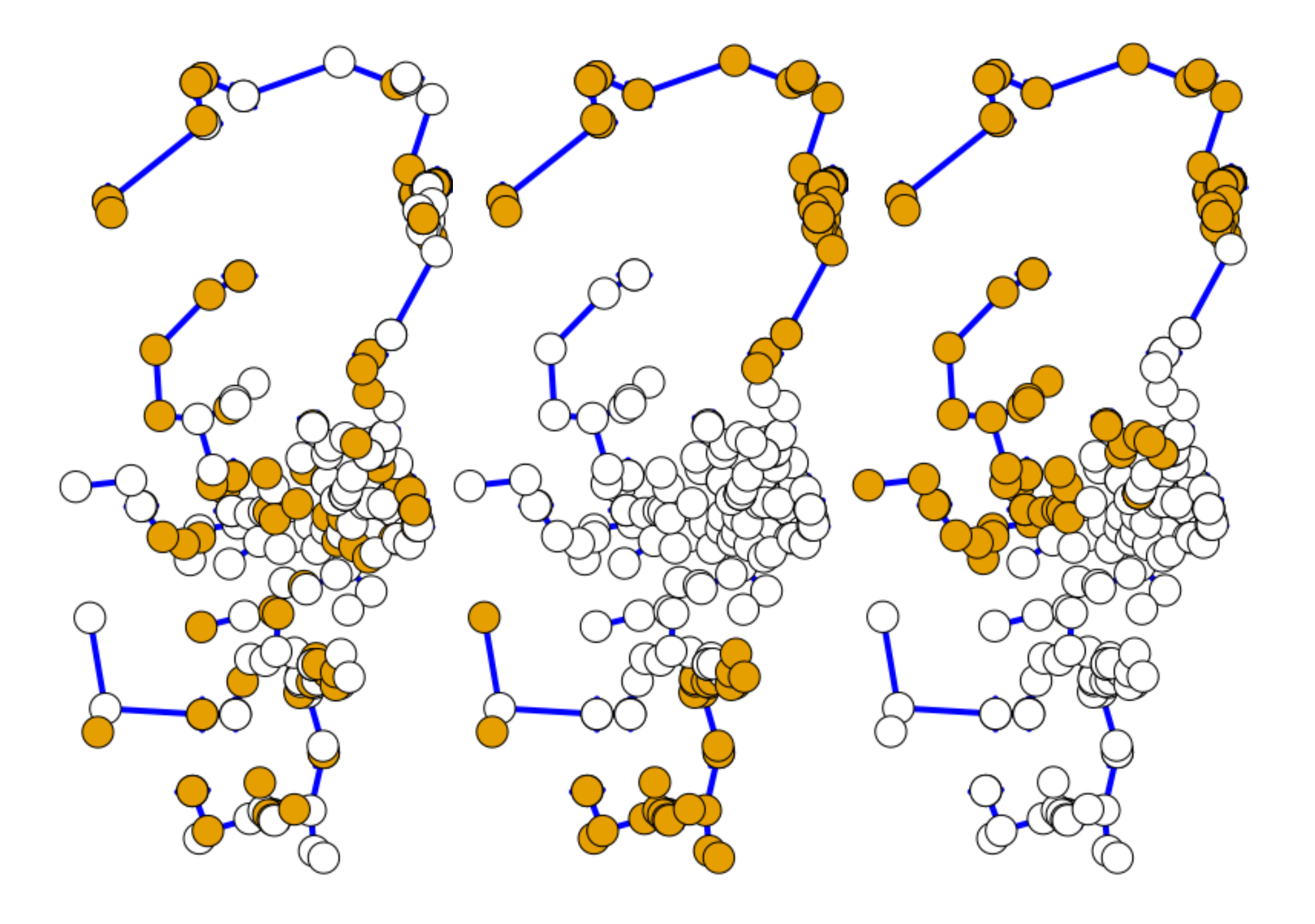}
\caption{Examples of system disruption types, orange vertices are operational, white have failed. Left is a network after \emph{random} failures, center is a network after \emph{BFS} failures, and right is a network after \emph{DFS} failures. All three represent a failure of 60\% of the vertices in the system.}\label{fig:three_types}
\end{figure}

We aim to evaluate how asymmetry in the impact across networks influences measurements of system performance, and to do so we evaluate three methods of disruption generation. The first are random disruptions in the system in which each node has an independent and identical probability of failure. Random failures are representative of system aging or general degradation. The second and third methods are derived from search trees and generate disruptions in the graphs which are spatially connected. The second method (BFS) uses a \emph{breadth-first search} tree to create locally clustered distributions of failures around a randomly selected root node. The third method (DFS) uses a \emph{depth-first search} tree to create a connected cluster of failures propagating away from a randomly selected root node and progressing away from the root to maximal length. Examples of each failure generation method are seen in Figure \ref{fig:three_types}. The three disruption types are selected to isolate the impact of the spatial distribution of failures on the interdependent systems. In this way, we can evaluate how a disruption which induces failures asymmetric to the two systems, impacts overall system performance and measurements of system resilience. The three failure generation methods listed here are all used to generate the set of initial failures $f^f_{g1}$ and $f^f_{g2}$.

\subsection{Recovery}
After the initial failures in the system are generated, the system is repaired sequentially. $r_{g1}^n$ is the $n^\text{th}$ node repaired in the graph $g_1$ such that $r_{g1}^n \in \{ f_{g1}^f \backslash r_{g1}^1, \dots, r_{g1}^{n-1} \} $\footnote{ In this notation, $\backslash$ indicates the removal of the vertices $r_{g1}^1, \dots, r_{g1}^{n-1}$ from the set $f_{g1}^f$}. That is, the eligible nodes for repair in $g_1$ at step $n$ are those which are have failed directly ($f_{g1}^f$) but have not yet been repaired ($r_{g1}^1, \dots, r_{g1}^{n-1}$). At step $n$, $r_{g1}$ and $r_{g2}$ are selected such to maximize the total system performance improvement. Total system performance is simply the sum of each individual network's performance. After $r_{g1}$ and $r_{g2}$ are repaired, any dependent failures (\ie elements of $f^d_{g1} \cup f^d_{g2}$ connected to $r_{g1}$ and $r_{g2}$ via $D$) are also repaired.  In this way, we can differentiate between repairs of directly failed elements of the systems and repair of elements which have only failed because of their dependency. The recovery and repair procedure is continued until both networks are fully operational.

\subsection{Data and parameters}
The networks in our model are based on publicly available electric power distribution grid location and the natural gas pipeline layout of Mobile County, Alabama. The electric power distribution system contains 223 vertices and 222 edges, while the natural gas system contains approximately 25 vertices and 35 edges \footnote{Estimates of the gas pipeline network are take from public-level aggregated pipeline locations available through the National Pipeline Management System}. In this analysis, the failure and recovery of the system is simulated on randomly generated \er graphs of equivalent degree. The degree of interdependence is estimated based on the physical proximity of nodes in the system which results $D$ having a matrix density of 0.01. Both networks are assigned a failure generation method (Random, BFS, DFS), and failures are generated such that 10, 20, 60, and 90\% of the components fail -holding the size constant in each replication. Every parameter combination (failure size, generation method in $g_1$, and generation method in $g_2$) is simulated 250 times, wherein each trial generates the failures randomly from one of the three methods.

Network performance is measured after the initial failures and is recorded throughout the recovery process. System performance is measured as the \emph{global efficiency} of each network. Global efficiency is defined as 

\[
\text{Eff}(G)=\frac{1}{n(n-1)}\sum_{i<j \in G} \frac{1}{d(i,j)}
\]

where $d(i,j)$ is the distance between vertex pair $i$ and $j$. Network efficiency as a concept was introduced by Latora (2001)  as a measure of how efficiently a network exchanges information \cite{latora2001}. It has been evaluated in the context of power system resilience evaluation \cite{larocca} and used as a proxy for network performance \cite{sun2017,winklerjames2011}.

\section{Results}

\subsection{Spatial differences in initial disruption}

\begin{figure}[htb]
\centering
\includegraphics[width=5in]{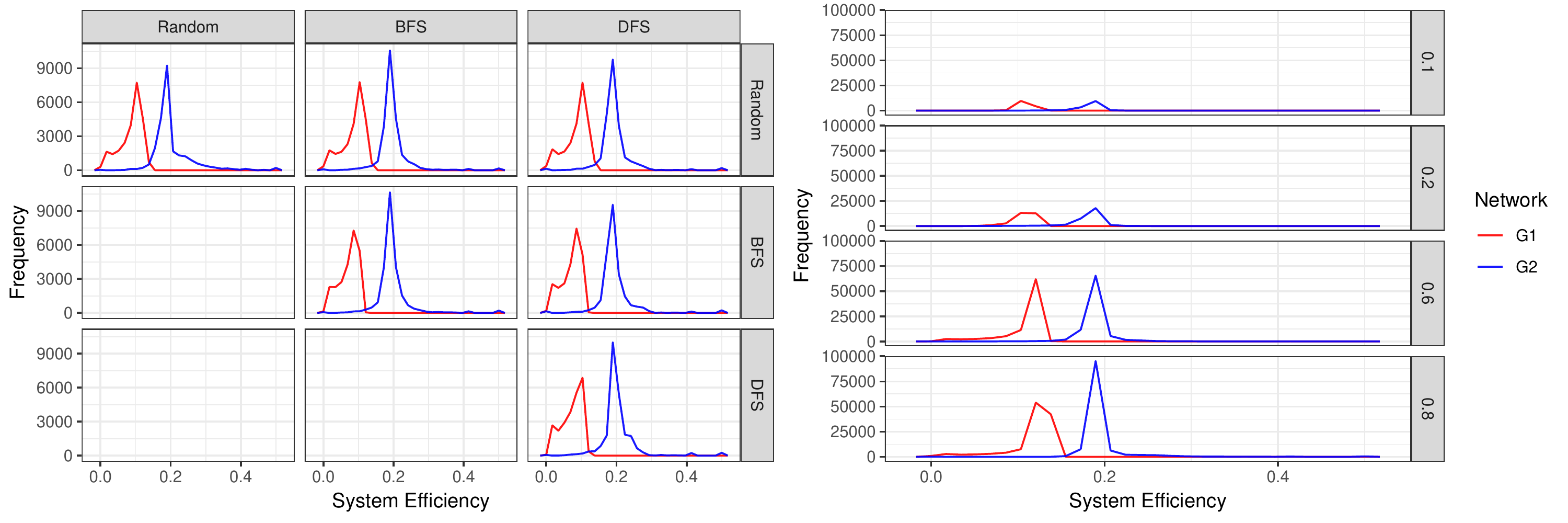}
\caption{Differences in initial disruptions. Left are density plots of both graphs subset by the corresponding distribution of failures in $g_1$ (rows) and $g_2$ (columns). Right shows the change in performance measure as a function of failure size (rows).}\label{fig1}
\end{figure}

Immediately after the failures in the system have completed propagating, we measure the efficiency of both systems in each replication. The distribution of the efficiency is listed for both systems in Table \ref{tab1} and density plots of the respective efficiency can be seen in Figure \ref{fig1}. At a fixed size of disruption, changing the spatial distribution of outages (or the \textit{shape} of the outages) in either network impacts the overall system performance for both networks. Table \ref{tab2} shows the results of two-sample Kolmogorov-Smirnov tests comparing the distributions of network efficiency for both systems after failures induced by different methods. The results of the KS tests show that there is a statistically significant difference in the performance of $g_1$ and $g_2$ when changing the spatial distribution of either network away from a random field of outages. 

\begin{table}[thb]
\centering
\caption{Efficiency of networks in different under different failure regimes}\label{tab1}
\begin{tabular}{ll|lll|lll}
\hline
$g_1$ Distribution & $g_2$ Distribution& Min, $g_1$ & Mean, $g_1$ & Max, $g_1$ & Min, $g_2$ & Mean, $g_2$ & Max, $g_2$ \\ \hline
Random   & Random   & 0.004       & 0.108    & 0.132   & 0       & 0.189    & 0.5   \\
Random   & BFS      & 0.005       & 0.108    & 0.132   & 0       & 0.188    & 0.5     \\
Random   & DFS      & 0.006       & 0.108    & 0.132   & 0       & 0.186    & 0.5     \\
BFS      & BFS      & 0.007       & 0.097    & 0.119   & 0       & 0.188    & 0.5     \\
BFS      & DFS      & 0.007       & 0.096    & 0.119   & 0       & 0.187    & 0.5     \\
DFS      & DFS      & 0.007       & 0.097    & 0.117   & 0       & 0.193    & 0.5    
\end{tabular}
\end{table}

\begin{table}[htb]
\centering
\caption{Comparison of initial disruption types. P-values taken from 2-sample Kolmogorov-Smirnov tests. Values of $2.2 \text{x} 10^{-16}$represent a p-value smaller than the numerical precision of R }\label{tab2}
%\vspace{-0.7cm}%Workaround to be conform with the .doc style. Only for table captions.
\begin{tabular}{l|cc}
\hline
Statistical test                        & P-value for difference in $g_1$ & P-value for difference in $g_2$ \\
\hline
Random-Random vs Random-BFS & 0.00089                         & $2.2 \text{x} 10^{-16}$                        \\
Random-Random vs Random-DFS &$ 1.576 \text{x}10^{-9}$                        & $2.2 \text{x} 10^{-16}$                    \\
Random-Random vs BFS-BFS    & $2.2 \text{x} 10^{-16}$                      & $2.2 \text{x} 10^{-16}$                       \\
Random-Random vs BFS-DFS    & $2.2 \text{x} 10^{-16}$                       & $2.2 \text{x} 10^{-16}$                       \\
Random-Random vs DFS-DFS    & $2.2 \text{x} 10^{-16}$                    & $2.2 \text{x} 10^{-16}$                       
\end{tabular}
\end{table}

\subsection{Changes in recovery of systems}

As the network is repaired, we measure changes in the performance of both systems. Figure \ref{fig3} shows the recovery of the systems as they are repaired subset by the distribution of outages in $g_1$ and $g_2$. Similar to in the previous section, changes in the distribution of failures in either system induce changes in the overall recovery of the system. Additionally, changes in the distributions of outages - \ie from random to a spatially-constrained outage- effect the variability of observations, with random outages exhibiting the highest variability among outage types.

% conclusion: variability changes, and values change. Providing evicence that (1) there might be a 

\begin{figure}[htb]
\centering
\includegraphics[width=4in]{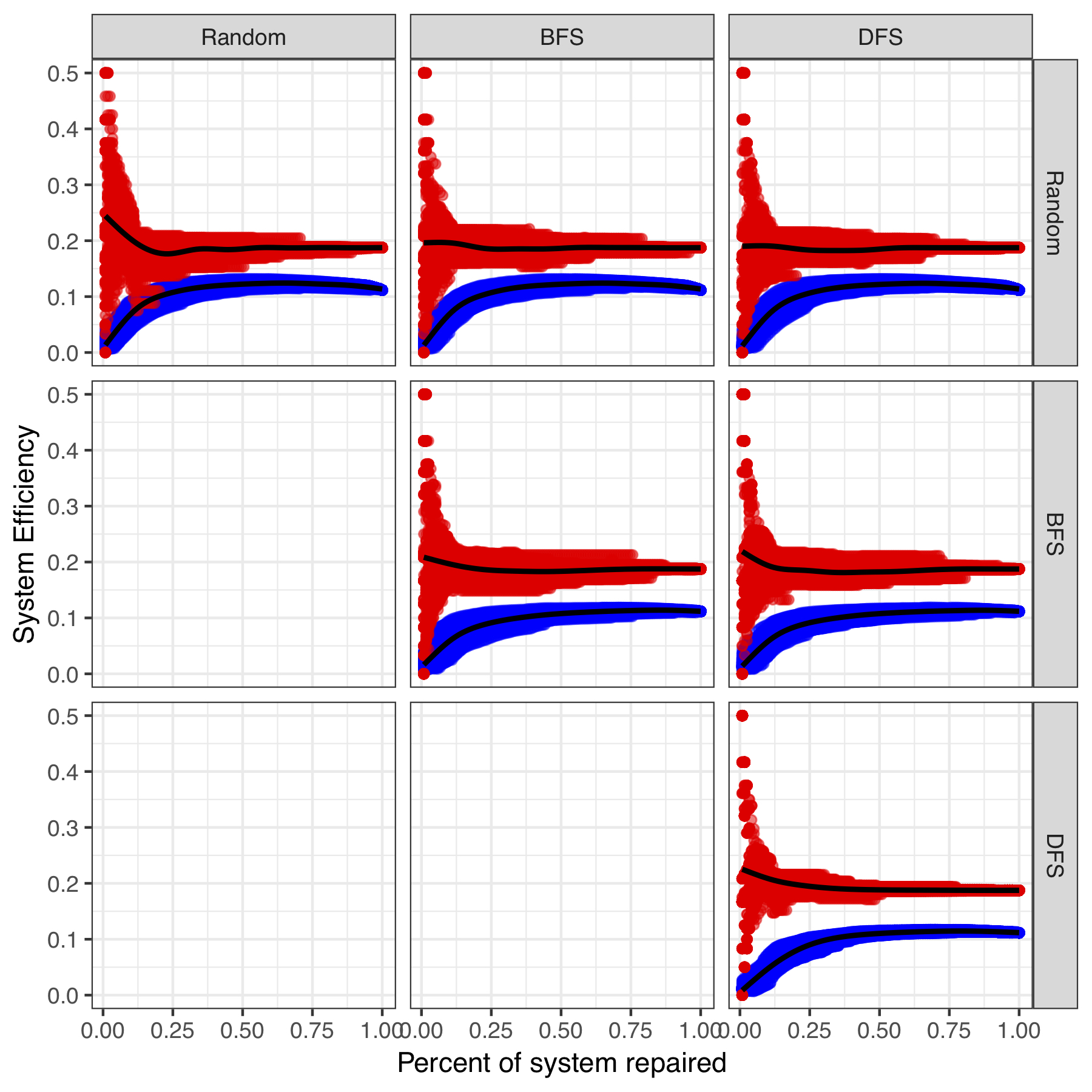}
\caption{Recovery of system over time subset by the distribution of failures in $g_1$ (rows) and $g_2$ (columns). Red points are the system efficiency for $g_1$, and blue are the efficiency of $g_2$. Black lines indicate the mean of all replications at a given percentage of the system repaired. }\label{fig2}
\end{figure}

In each simulation replication, the time is measured after failure until the system is exhibiting full performance. Because of redundancies in the network, it is frequently the case that the the system is fully operational \textit{prior to} all elements being repaired. The Time to Repair (TTR) in this case is measured as the first time a system is performing optimally in a given replication. Figure \ref{fig3} shows the TTR broken down by failure size and disruption type. As expected, larger failure sizes have higher TTR -corresponding to a longer time to repair. However changes in the time to repair can be observed when the distribution of failures is altered in either network. 

\begin{figure}[!htb]
\centering
\includegraphics[width=5in]{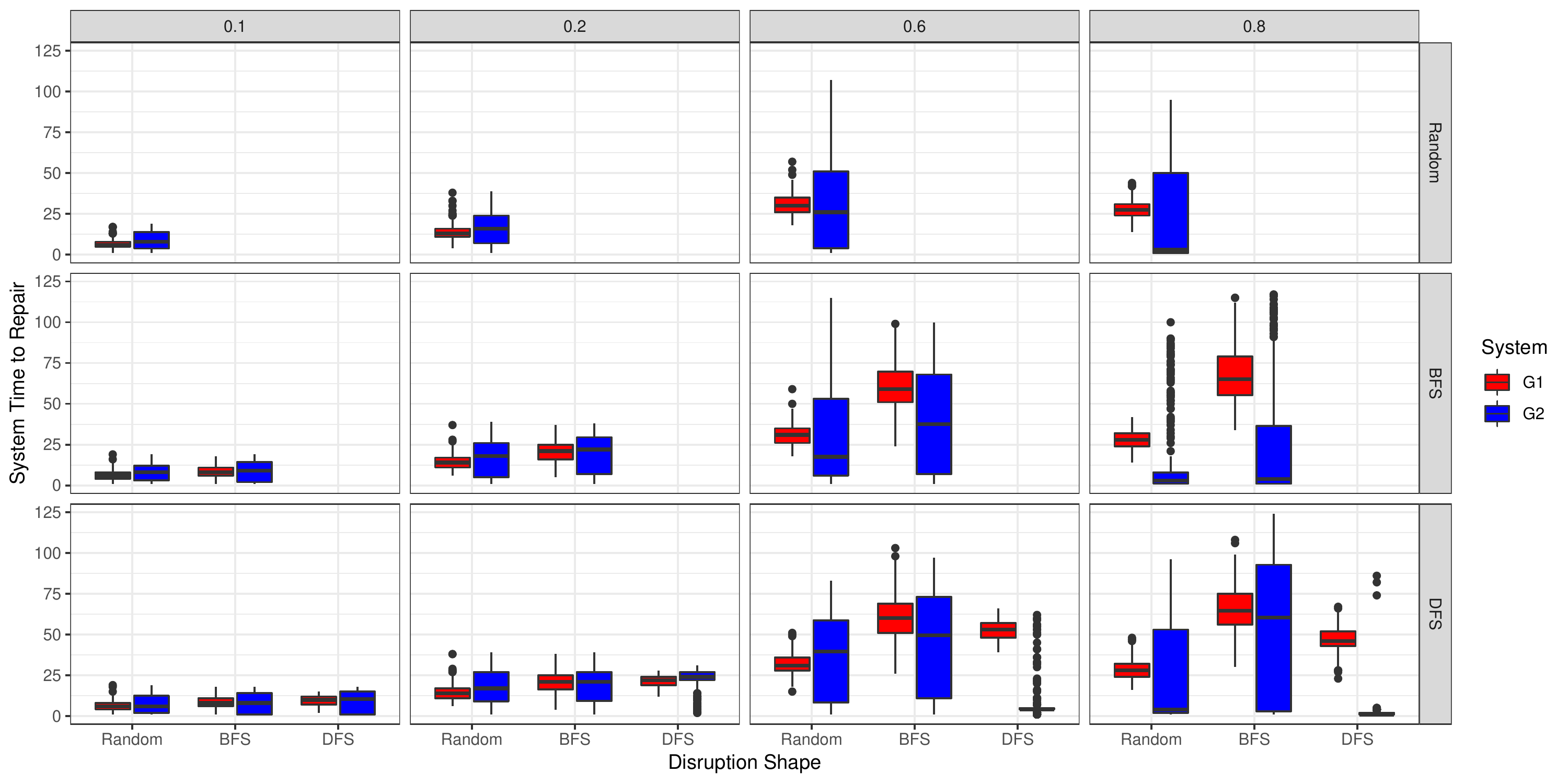}
\caption{Differences in time to repair (TTR) by failure method subset by the size of failures (columns) and the distribution of failues in $g_1$ (rows). Each individual plot shows the difference in TTR for $g_1$ (red) and $g_2$ (blue) for changes in the distribution of failures in $g_2$ (sub-columns).}\label{fig3}
\end{figure}

\section{Conclusion}

In this work, we construct a simulation of interdependent networks, representing coupled infrastructure, which are subsequently disrupted and repaired. We hypothesize that a major hazard which disrupts interdependent systems will impact the constituent systems asymmetrically, inducing different magnitudes of failures and different spatial distributions of failures in each system. Via leveraging a rigorous simulation methodology to test our hypothesis, we provide evidence  that the differences in the system performance can be observed when the spatial distribution of failures is changed; this is done while also controlling for the effect of the disruption size. The spatial distribution of the failures additionally changes the recovery of the system- measured by the time to system repair and functional recovery form. Consideration the impact of a disturbance on each network within an interdependent system can provide better assessments of infrastructure and system risk and resilience.

\bibliographystyle{ieeetr} 
{\footnotesize
\bibliography{bib}}

\end{document}